\UseRawInputEncoding

\documentclass{ijcaArticle}
\ijcaVolume{174}
\ijcaNumber{30}
\ijcaYear{2021}
\ijcaMonth{April}

\usepackage{graphicx}
\usepackage{blindtext}
\usepackage{float}
\usepackage{svg}
\usepackage{mathrsfs}
\usepackage{environ}
\usepackage{amsmath}
\graphicspath{ {./images/} }

\NewEnviron{mymath}{
	\begin{math}
		\scalebox{1.3}{$\BODY$}
	\end{math}
}

\begin{document}

\title{LiveStyle - An Application to Transfer Artistic Styles}

\author{
	\large Amogh G. Warkhandkar \\[-3pt]
	\normalsize Vidyalankar Institute of Technology \\[-3pt]
	\normalsize Mumbai, \\[-3pt]
	\normalsize Maharashtra, India \\[-3pt]
	\normalsize amoghw2025@gmail.com \\[-3pt]
	\and
	\large Omkar B. Bhambure \\[-3pt]
	\normalsize Vidyalankar Institute of Technology \\[-3pt]
	\normalsize Mumbai, \\[-3pt]
	\normalsize Maharashtra, India \\[-3pt]
	\normalsize omieblablablu2996@gmail.com \\[-3pt]
}

\terms{Artificial Intelligence, Machine Learning, Deep Learning, Image Processing, Computer Vision, Software Engineering}


\maketitle

\begin{abstract}
	Art is a variety of human activities that include the production of visual, auditory, or performing objects that express the creativity, creative concepts, or technological abilities of the artist, intended primarily for their beauty or emotional power to be appreciated. The renaissance of historic and forgotten art has been made possible by modern developments in Artificial Intelligence. Techniques for Computer Vision have long been related to such arts. Style Transfer using Neural Networks refers to optimization techniques, where a content image and a style image are taken and blended such that it feels like the content image is reconstructed in the style image color palette. This paper implements the Style Transfer using three different Neural Networks in form of an application that is accessible to the general population thereby reviving interest in lost art styles.
\end{abstract}

\section{Introduction}

Art enables humans to express their creativity in various forms like visual, auditory, sensory, etc. Photos are an essential part of modern human life. The advancement in computers has paved a way for humans to connect to their roots. This application \footnote{Source code available at: \href{https://github.com/amogh-w/LiveStyle}{LiveStyle}.} aims to explore these lost arts and re-introduce the styles in modern human civilization. Users can upload their photos along with a painting and have their photo brought back to life as if the original artist has created it. Image Style Transfer is a task that aims to render the content of one image with the style of another, which is relevant and fascinating for both practical and scientific reasons. In image processing applications such as mobile camera filters and artistic image generation, Style Transfer techniques can be widely used. Furthermore, the study of Style Transfer also shows the intrinsic features of photographs. Style Transfer is daunting as it is difficult to achieve a clear distinction of styles concerning the content image as art is subjective.

\begin{figure}
	\includegraphics[width=\linewidth]{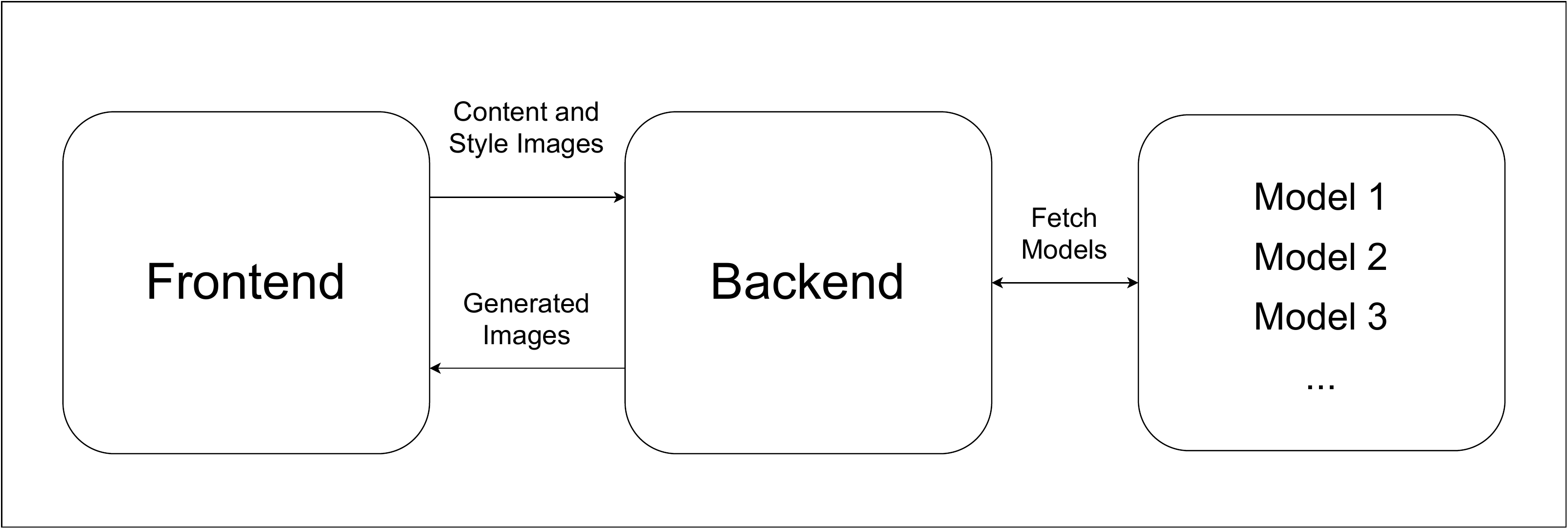}
	\caption{Flow Diagram}
	\label{fig:flowdiagram}
\end{figure}

The flow, Fig. \ref{fig:flowdiagram} of the application is elementary. The application accepts content and style images as input from the user. To be able to upload images smoothly and build a progressive web application, it is built using React \cite{react} and MaterialUI \cite{materialui} library. The backend library used in this application is FastAPI \cite{fastapi}. For version control, Git \cite{git} was utilized, which gave a contribution to Continuous Integration/Continuous Delivery (CI/CD). To provide the application in a portable, light-weighted containerized environment, Docker \cite{docker} was used. Docker handles infrastructure management and significantly reduces the delay between writing code and running it in production. The interface connecting the frontend and the backend must be compatible and should be able to support Dynamic Rate Changes and Load balancing features such as to not crash under heavy load.

\section{Implementation}

\subsection{React (Frontend)}

The frontend library used in this application is React. For building user interfaces, React is a declarative, powerful, and scalable JavaScript library. React is an open-source, component-based frontend library that is responsible for the application's view layer. React uses a declarative model that makes thinking about the application simpler and strives to be both productive and versatile. It designs simple views for each state in your application, and when your data changes, React can efficiently update and make just the right component. The code is made more predictable and easier to debug by the declarative view. A React application consists of several modules, each responsible for rendering a small piece of HTML that can be reused. To allow complex applications to be built out of simple building blocks, components can be nested within other components.
For routing and managing the pages, the \textit{react-router} package was used. We used the component-based structure to store our files for easier future upgrades and faster code reviews. Each file in the components folder contains JavaScript code for rendering the elements on the website.

\textit{User Interface} and \textit{Usability} is an important concept while building the Frontend of an application. This requires proper planning and creative thinking. The software which is being used to create these User Interface elements should be carefully examined to suit the needs of the project.

\subsection{FastAPI (Backend)}

The backend library used in this application is FastAPI. FastAPI is a software framework for the Python creation of web applications. For validating, serializing, and deserializing data, FastAPI refers to \textit{Pydantic} and type hints and automatically generates OpenAPI documents. Asynchronous programming is fully supported and can it can run with Uvicorn \cite{uvicorn} and Gunicorn \cite{gunicorn}. Backend should be able to handle requests from all devices without bottlenecking.

The main.py file contains the logic for the backend. In this file, the libraries are imported first, followed by defining the routes where the frontend will send the content and style images. The server will process the images and return the output image to the origin of the request.

\subsection{TensorFlow (Deep Learning Library)}

TensorFlow \cite{tensorflow} is a free and open-source software library for machine learning. It contains a comprehensive, flexible ecosystem of tools, libraries, and community resources that enable researchers to advance the state of the art in ML and that developers can easily create and deploy ML-based applications. Machine Learning models can be built easily using higher-level APIs like Keras. Training and Evaluation are supported in the cloud, on-premise, in the browser, or on the device, regardless of the language being used.
The models were trained on an \textit{NVIDIA GTX 1660 Super} with 6GB of GDDR6 Memory with \textit{CUDA} support. The models were then exported as h5 files, which were then loaded by the server.
Models used for the Style Transfer must be deployed on the backend which can support high-performance computing and processing.

\subsection{Model 1: VGG19}

VGG-19 \cite{simonyan2015deep} is a 19 layer deep convolution neural network. In the implementation by Gatys et al. \cite{DBLP:journals/corr/GatysEB15a} a pre-trained version of the network is loaded which is trained on more than a million images \footnote{ImageNet dataset available here: \href{http://www.image-net.org/}{ImageNet}.}. This model can classify images into 1000 object categories. This model has learned rich feature representations for a wide range of images. Over the iterations, Content Loss and Style Loss is calculated at every iteration. These are combined in the equation of Total Loss which the network tries to minimize.

The functions for Style and Content Loss are as follows -

\vspace{3mm}

\begin{mymath}
	\mathscr{L}_{style}(\overrightarrow{a},\overrightarrow{x} )  = \sum_{l = 0}^{L} w_{l}E_{l}
\end{mymath}

\vspace{3mm}

\begin{mymath}
	\mathscr{L}_{content}(\overrightarrow{p},\overrightarrow{x},l)  = \frac{1}{2} \sum_{i, j}^{} (F_{ij}^{l} - P_{ij}^{l})^{2}
\end{mymath}

\subsection{Model 2: Inception-v3 + MobileNet-v2}

Inception-v3 \cite{DBLP:journals/corr/SzegedyVISW15} is a 48 layer deep convolution neural network. This model can also classify images into 1000 object categories. This model is made up of symmetric and asymmetric building blocks. In the implementation by Ghiasi et al. \cite{DBLP:journals/corr/GhiasiLKDS17} used a pre-trained MobileNet-v2 \cite{DBLP:journals/corr/abs-1801-04381} model to distill the trained Inception-v3 model which allows the model to run on low computing devices.

AST utilizes a Style Transfer network that can map from content image to stylized image. The specific style is chosen by providing the transfer network with a style-specific set of normalization parameters. These parameters are generated for any arbitrary painting via a separate neural network called the style prediction network. The style-specific parameters can be taken as a latent space of style, and we can ``combine" different styles by taking the weighted average between style representations of multiple styles. AST allows interpolation between the style representation of arbitrary styles, and so lets us do something like control the stylization strength. This is done by calculating the style representation of both the content image and the style image and taking their weighted average.

The functions for Style and Content loss are as follows -

\vspace{3mm}

\begin{mymath}
	\mathscr{L}_{style}(x, s)  = \sum_{i\epsilon s}^{} \frac{1}{n_{i}} \| G[f_{i}(x)] - G[f_{i}(s)] \| ^2_{F}
\end{mymath}

\vspace{3mm}

\begin{mymath}
	\mathscr{L}_{content}(x, c)  = \sum_{j\epsilon c}^{} \frac{1}{n_{j}} \| f_{i}(j)- f_{j}(s)\| ^2_{2}
\end{mymath}

\subsection{Model 3: CycleGAN}

The generator tries to produce samples from the desired distribution and the discriminator tries to predict if the sample is from the actual distribution or produced by the generator. The benefit of the CycleGAN \cite{DBLP:journals/corr/ZhuPIE17} model is that it can be trained without paired examples. It does not require examples of photographs before and after the translation to train the model, e.g. image of the dog and a painting of the same dog in an artist’s style.  Instead, the model can use a collection of images from each domain and extract and harness the underlying style of images in the collection to perform the translation.

The loss function of this model is composed of two parts: the Adversarial Loss L(adv) and the Content Loss L(con). The content loss is used to constrain semantic content between the input images and output images. The hyperparameter lambda is used to control how much input image content is retained in the image stylization. Its value varies with the specific Style Transfer.

This model tries to convert from Domain X to Domain Y using adversarial loss. As this mapping is highly under-constrained, it is coupled with an inverse mapping and a cycle consistency loss is introduced.

The loss function in this implementation is as follows -

\vspace{3mm}

\begin{mymath}
	L(G, D) = L_{adv}(G,D) + \lambda L_{con}(G,D)
\end{mymath}

\section{Observations}

\subsection{User Interface}

\begin{figure}[H]
	\includegraphics[width=\linewidth]{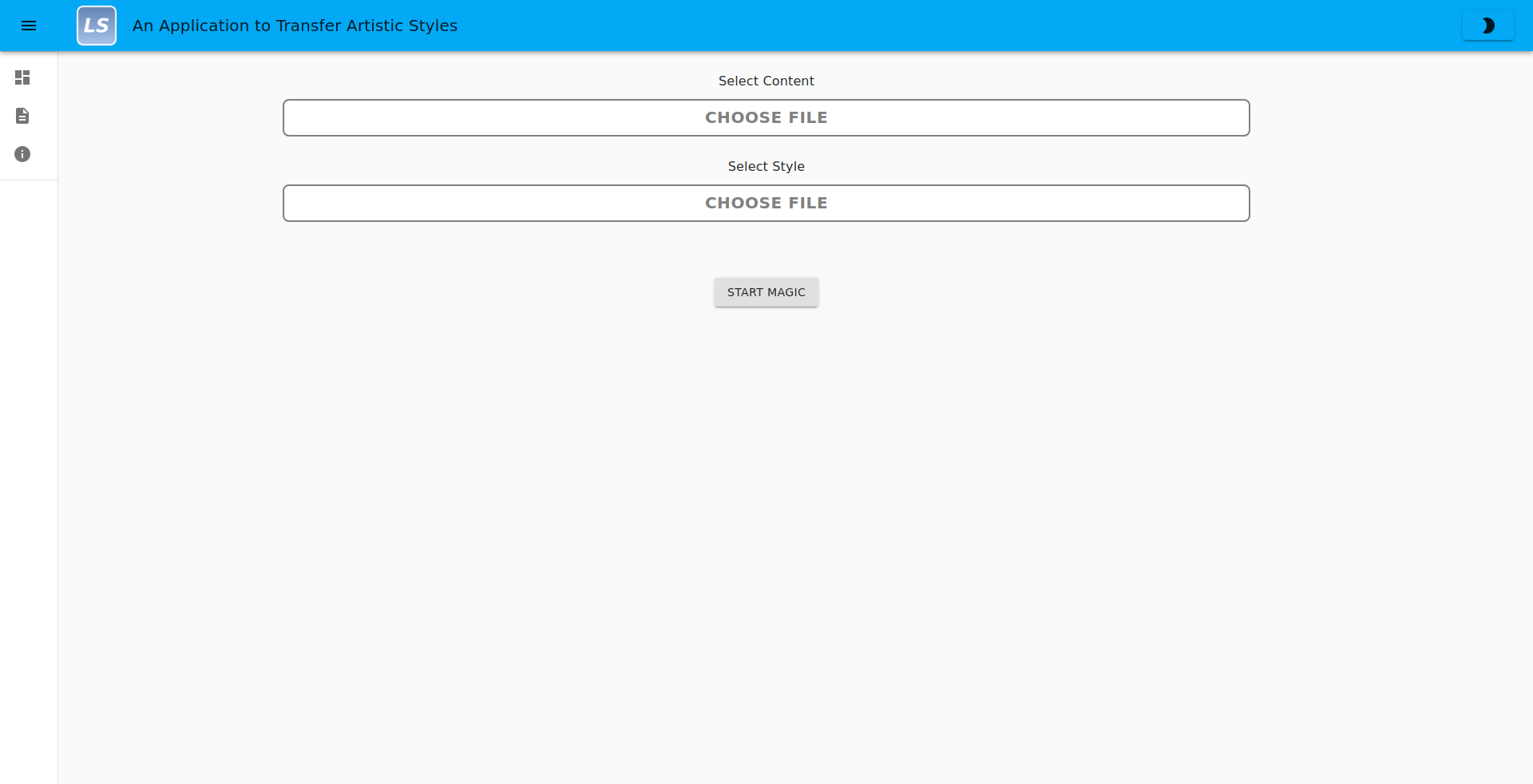}
	\caption{Dashboard}
	\label{fig:dashboard}
\end{figure}

The Dashboard Page, Fig. \ref{fig:dashboard} features two buttons that the user can click. This opens the native file picker of the Operating System being used. After selecting the content and style image the user can click the ``Start Magic" Button. This sends a GET request to the backend. The server processes the two images and returns the output of the models. This output is then displayed on the frontend.

\begin{figure}[H]
	\includegraphics[width=\linewidth]{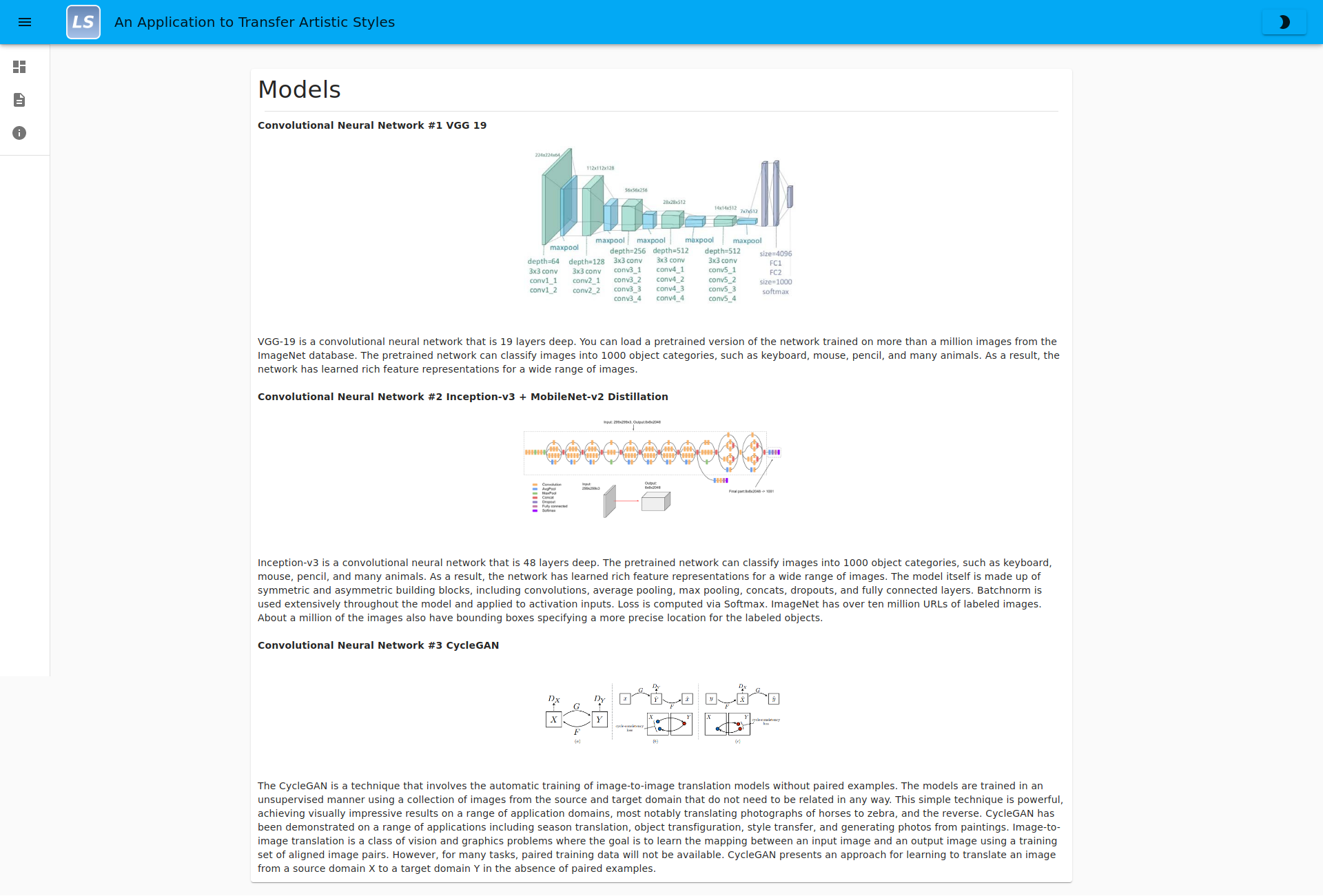}
	\caption{Resources}
	\label{fig:resources}
\end{figure}

The Resources Page, Fig. \ref{fig:resources} displays the details of the models being used in this application.

\begin{figure}[H]
	\includegraphics[width=\linewidth]{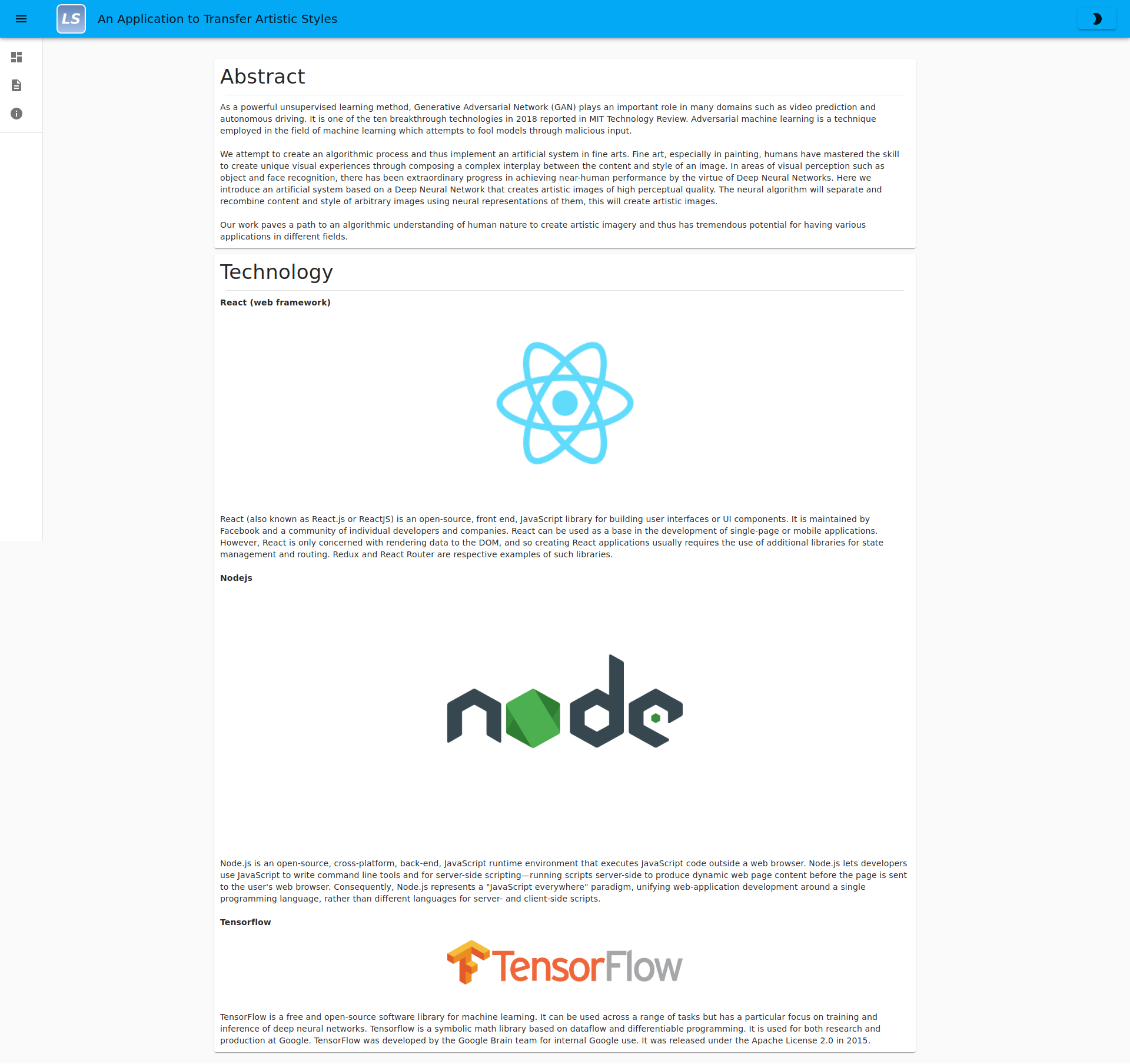}
	\caption{About}
	\label{fig:about}
\end{figure}

The About Page, Fig. \ref{fig:about} contains details of the application and the technology stack being used.

\subsection{Generated Images}

\begin{figure}[H]
	\includegraphics[width=\linewidth]{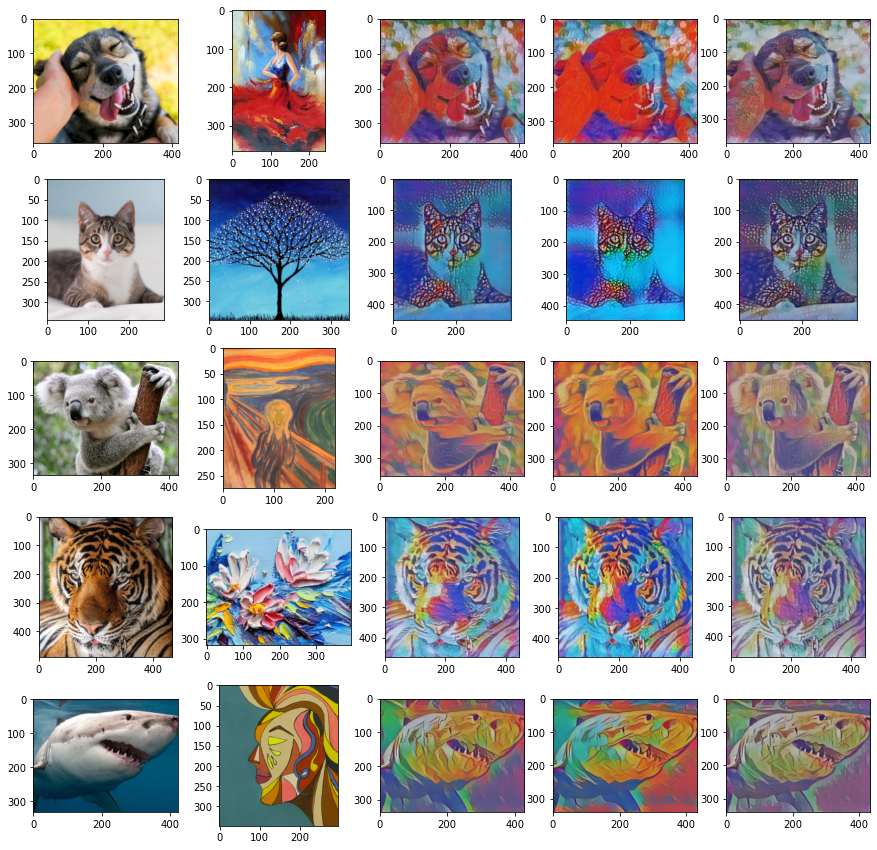}
	\caption{Generated Images}
	\label{fig:generated}
\end{figure}

Fig. \ref{fig:generated} indicates the style and content image used for comparative and qualitative analysis of the models. The style images are the paintings of various artists. For the demonstration of the working of the application, images of animals were taken as content images. This figure displays the output images generated by the models for comparison. As there is no objective metric to judge an artwork, one can only compare the output image with the input images. Table \ref{tab:subjective_ranking} shows the quality of generated images in the form of rankings with rank 1 being the best in the mentioned comparison metric.

\begin{table}[H]
	\tbl{Subjective Ranking of Generated Images}{
		\begin{tabular}{|l|l|l|l|l|}
			\hline
			Model        & Sharpness & Noise & Color Accuracy & Distortion \\
			\hline
			VGG19        & 2         & 2     & 3              & 1          \\
			\hline
			Inception-v3 & 1         & 1     & 2              & 3          \\
			\hline
			CycleGAN     & 3         & 3     & 1              & 2          \\
			\hline
		\end{tabular}}
	\label{tab:subjective_ranking}
\end{table}

\section{Conclusions and Future Work}

React allows us to build the frontend and user interface elements easily. FastAPI helps us construct an endpoint where it is possible to host TensorFlow models. Git enables the management of files and keeping track of software versions. Docker is a great tool for containerizing the dependencies needed to compile, build, run, and test isolated program instances.

Thus, for future work, it would be interesting to find efficient methods of performing Style Transfer that can provide faster results. Users would be offered more customization options for the generated image. The Style Transfer application can be also extended to support live video feed.

\nocite{*}
\bibliographystyle{ijcaArticle}
\bibliography{References}

\end{document}